\documentclass[pdflatex,prb,twocolumn,showpacs]{revtex4}
\usepackage{graphicx}% Include figure files
\usepackage{dcolumn}% Align table columns on decimal point
\usepackage{bm}% bold math
\usepackage{amsmath,epsf}
\usepackage[colorlinks,breaklinks,citecolor=red,pagebackref]{hyperref}

\begin{document}

\author{Dipta Bhanu Ghosh and Stefano de Gironcoli}
\affiliation{INFM-CNR DEMOCRITOS National Simulation Center, Trieste 34014,
Italy \\ SISSA -- Scuola Internazionale Superiore di Studi Avanzati ,
Trieste 34014, Italy }

\title{A new look on the nature of high-spin to low-spin transition
in Fe$_{2}$O$_{3}$}

\date{\today}% It is always \today, today,

\begin{abstract} 
Iron sesquioxide (Fe$_{2}$O$_{3}$) displays pressure and temperature
induced spin and structural transitions. Our calculations show that,
density functional theory (DFT), in the generalized gradient approximation
(GGA) scheme, is capable of capturing both the transitions. The ambient
pressure corundum type phase (hematite or $\alpha$--Fe$_{2}$O$_{3}$), having $R_{\bar 3 c}$
symmetry, gets distorted by the application of pressure and transforms to
a distorted corundum type or Rh$_{2}$O$_{3}$(II) phase with $Pbcn$ symmetry,
in agreement with recent experiments. GGA + $U$ calculations show the same
trend but shift the transition pressures to higher values. Experimentally,
the onset of the structural transition begins in the vicinity of the spin
transition pressure and whether the system undergoes spin transition in
the corundum type (HP1) or in the Rh$_{2}$O$_{3}$(II) type (HP2) phase, is still a
controversial issue.  With a relatively simple, but general, octahedral
structural parameter, $V_{oct}$ (the octahedral volume around iron ions),
we show that in order to acquire a low spin (LS) state from  a high spin
(HS) one, the system does not necessarily need to change the crystal
structure. Rather, the spin transition is a phenomenon that concerns
the cation octahedra and the spin state of the system depends mainly on
the value of $V_{oct}$, which is governed by two distinct equations of
state, separated by a well defined volume gap, for the HS and LS states
respectively. Analysis of the results on the basis of octahedral volume
allows to sum up and bridge the gap between two experimental results
and thus provides a better description of the system in the region
of interest.

%Concurrence of pressure and/or temperature induced spin and structural transitions in Fe$_{2}$O$_{3}$ makes it difficult to understand experimentally their relationship. By extensive DFT calculations on several candidate phases we have been able to reproduce and separate the two transitions. Our analysis shows that spin transition is present in all structures and ruled by a simple local structural parameter, the octahedral volume around Fe ions. Moreover it suggests that at intermediate pressures the system is in a low-spin paramagnetic phase, thus clarifying some experimental controversy.
\end{abstract}
%\pacs{91.60.Gf,71.15.Mb,71.30.+h,75.30.Kz}

\maketitle

\section{INTRODUCTION}

Spin crossover or electronic transition is a phenomenon commonly
encountered among (transition) metal complexes, in particular
under octahedral coordination \cite{Takemoto-ic73,Gutlich-sb81, 
Konig-pic87,Konno-bcsj91,Moliner-ica99, Guionneau-jmc99, 
Marchivie-ac03}. Under the influence of the ligand field the metal
$d$ orbitals are split into two sets: a $t_{2g}$ level with a 3-fold
degeneracy and a 2-fold degenerate $e_{g}$ level. The closer the ligand
is to the metal ion, the greater is the influence of the ligand-field
strength. When this ligand-field energy dominates over electron-pairing
or exchange energy the metal ion adopts the low-spin configuration. A
high-spin configuration is instead favored when the exchange energy is
greater than the ligand-field energy.

Hematite, $\alpha$-Fe$_{2}$O$_{3}$, is a rhombohedrally
structured wide-gap antiferromagnetic insulator\cite{Hubbard64}
at ambient conditions. It becomes weakly ferromagnetic between
Morin\cite{Morin-pr50} temperature, T$_{M}$ (=260 K), and N\'eel
temperature, T$_{N}$ (= 955 K), as a result of the canting of the spins
of the two sublattices \cite{Searle-prb70, Levinson-prb71, Chow-prb74,
Sandratskii-epl96}. In this structure, the Fe$^{3+}$ ions, sitting in
octahedral cages, display a high spin (HS) to low spin (LS) transition
under external influences such as, temperature, pressure, etc. Sequential
or concomitant pressure and/or temperature induced structural phase
transitions, in addition to the spin one, increase the complexity of
the system, making its in-depth understanding a challenge.

Significant continuous attention by the scientific community,
for more than forty years, has brought in many intuitive ideas and
counter-ideas\cite{Reid-jgr69,Mcqueen-gsa66,Shannon-jssc70, Yagi-capj82}.
The existence of residual magnetism after a volume collapse of $\sim$
10 \% and concurrent onset of the structural phase transition led to the
issue of {\it one}--type or {\it two}--type cationic picture for the high
pressure phase \cite{Suzuki-ktksp85, Syono-ssc84, Nasu-hi86, Olsen-ps91}.
To resolve this issue and correlate/separate the spin and structural
transitions a number of investigations have been performed in more
recent years. Of them, X-ray diffraction (XRD),  M\"ossbauer spectroscopy
(MS) at 300K and electrical resistivity measurements by Pasternak {\it
et.~al.}\cite{Pasternak-prl99} assigned a nonmagnetic distorted corundum
or Rh$_{2}$O$_{3}$(II) type structure as the high pressure phase. The
intermediate region between the insulating and metallic regions in the
electrical resistivity data was qualitatively explained by the coexistence
of an insulating magnetic phase with a metallic nonmagnetic one.

Another XRD and X-ray emission spectroscopy (XES) experiment by
Badro {\it et.~al.}\cite{Badro-prl02} could separate the electronic
transition from the crystallographic one. From the XES data these authors
concluded that a high-spin state could be stabilized in the high-pressure
structural phase at low temperatures while at high temperatures a low-spin 
state was stable. 
This high temperature LS state was found to be still weakly paramagnetic as
evidenced by the presence of a satellite (K$\beta^{\prime}$) in the
XES spectra\cite{Badro-prl02}, in contrast with the non-magnetic state
suggested by the authors of Ref.\ \onlinecite{Pasternak-prl99}.
The structural transition shows a sluggish behavior, as reported 
in room temperature Raman spectroscopy studies by Shim and
Duffy\cite{Shim-am02} and angle-resolved XRD studies
by Rozenberg {\it et.\ al.}\cite{Rozenberg-prb02}. Moreover, 
high temperature heating might have a role to play 
in controlling the transition pressure as shown in the XRD 
experiments by Ono {\it et.\ al.}\cite{Ono-jpcs04,Ono-jpcm05}.

Hence, in spite of significant experimental efforts devoted to this
system, some still controversial and unresolved issues remain: (i)
how does the spin state of the system evolve with pressure ? (ii) what
is the correlation between the spin and the structural transition,
if there is any? In this study we will show that spin transition
is a general phenomenon, depending on a simple structural parameter,
the octahedral volume around Fe ions, and thus largely uncorrelated
to the structural transition occurring in the same pressure range.
The apparent contradiction between experimental results reported in Ref.\
\onlinecite{Pasternak-prl99} and \onlinecite{Badro-prl02} will also
be resolved.

\section{METHODS AND RESULTS}

Extensive first-principles density functional calculations
of the spin and structural transitions in Fe$_{2}$O$_{3}$ have been
performed. Variable-cell-shape molecular dynamics\cite{Wentzcovitch-prb91}
has been used for the full structural optimization at arbitrary pressures.
Eight-electron (3d$^{7}$4s$^{1}$) and six-electron (2s$^{2}$2p$^{4}$)
ultrasoft pseudopotentials \cite{Vanderbilt-prb90} were used to describe
Fe and O atoms\cite{pseudo}.  The electron--electron correlation
energy was treated within the PBE\cite{Perdew-prl96} generalized
gradient approximation (GGA). GGA+U calculations were performed to
supplement and compare with the results obtained within GGA. In these
cases, the U values have been calculated self-consistently within
the methodology described in Ref.[\onlinecite{Cococcioni-prb05}].
All calculations presented in this work have been performed with the
{\sc Quantum-ESPRESSO}\cite{QE} package.

\begin{figure}[!h]
\includegraphics[scale=0.24]{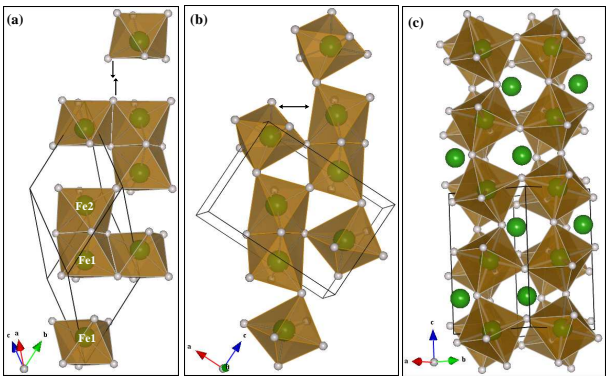} 
\caption{(Color online) Structures for the (a): corundum-type, (b): Rh$_{2}$O$_{3}$(II)-type
 and (c): {\it opv}-type Fe$_{2}$O$_{3}$.}
\label{structure}
\end{figure}

To understand the evolution of the spin and structural transition in the
system we have considered three possible structural phases, starting
from ambient to high pressure.  Fig. \ref{structure}(a) refer to the
rhombohedrally structured corundum-type Fe$_{2}$O$_{3}$, where the
four Fe atoms are arranged linearly along the rhombohedral axis in the
unit cell. The two possible candidate structures at elevated pressure
are Rh$_{2}$O$_{3}$(II)-type and orthorhombic-perovskite ({\it opv})
type Fe$_{2}$O$_{3}$, as shown respectively, in Fig. \ref{structure}(b)
and Fig. \ref{structure}(c) and contain four formula units per unit
cell. For each of the above mentioned structures there can be a few
anti-ferromagnetic configurations in addition to the ferromagnetic
one, depending on the spin orientations.

\par
\begin{figure}[!h]
\includegraphics[scale=0.40]{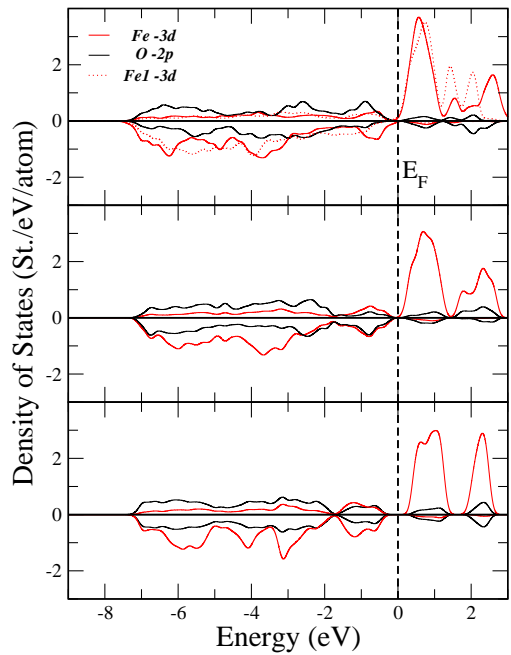} 
\caption{(Color online) Projected density of states calculated at 0 GPa within GGA. Lower panel: corundum-type. Middle panel: Rh$_{2}$O$_{3}$(II)-type and upper panel: {\it opv}-type Fe$_{2}$O$_{3}$. Note that in the upper panel the dotted lines ({\it Fe1 -3d}) correspond to Fe atoms with 8-fold coordination.}
\label{dos-hem-pbcn-pbnm}
\end{figure}

To start with we present the density of states (DOS) for
each of the abovesaid three structures for their most stable
ambient-pressure magnetic state. As can be seen from the lower panel
of Fig. \ref{dos-hem-pbcn-pbnm}, DFT at the GGA level correctly
describes the insulating nature of hematite. However, there is,
as typical of LDA/GGA calculations, an underestimation of the band
gap. Rh$_{2}$O$_{3}$(II)-type Fe$_2$O$_3$ also shows the same behavior,
but with a much diminished magnitude of the band gap, as shown in
the middle panel of Fig. \ref{dos-hem-pbcn-pbnm}. In both cases the
valence band top edge is equally populated by Fe--$3d$ and O--$2p$
states. However, DOS for the {\it opv}-type Fe$_{2}$O$_{3}$, shown in
the upper panel of Fig. \ref{dos-hem-pbcn-pbnm}, displays a very poor
metallic tendency with Fe--$3d$ dominated valence edge. This could be
due to the underestimation of correlation effects by GGA. However, this
issue is not relevant here since the latter structure is never the most
stable one in the pressure range of interest, as will be shown below.

\begin{figure}[!h]
\includegraphics[scale=0.32]{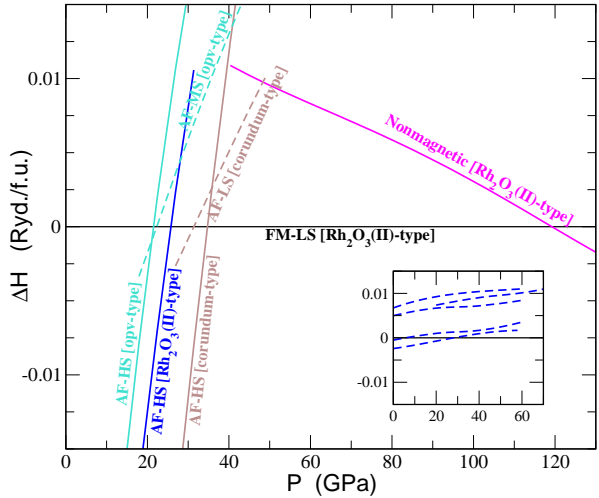} 
\caption{(Color online) Relative
enthalpy curves of Fe$_{2}$O$_{3}$ for corundum type,
Rh$_{2}$O$_{3}$(II) type and  orthorhombic perovskite-type ({\it opv})
structures in different magnetic configurations: ferromagnetic (FM),
antiferromagnetic (AF), with HS or LS, and non-magnetic. Inset shows 
different LS AF configurations for Rh$_{2}$O$_{3}$(II) type 
Fe$_{2}$O$_{3}$ with respect to the LS FM Rh$_{2}$O$_{3}$(II) 
type Fe$_{2}$O$_{3}$.}
\label{enthalpy-wrt-pbcn-fm-ls}
\end{figure}

The calculated relative enthalpy curves of Fe$_{2}$O$_{3}$
for corundum-type, Rh$_{2}$O$_{3}$(II)-type and {\it opv}-type
structures in different magnetic configurations are reported in Fig.\
\ref{enthalpy-wrt-pbcn-fm-ls}. As can be seen, the ambient-pressure
antiferromagnetic corundum-type phase, having a single kind of
polyhedral arrangement, is the most stable structure up to a pressure
of about 34 GPa. Starting from this pressure the distorted corundum or
Rh$_{2}$O$_{3}$(II)-type phase in the LS ferromagnetic configuration
is the thermodynamically stable phase.  
Several LS antiferromagnetic configurations are present a
few mRy above the FM ground state in the LS pressure range (see inset of 
Fig.\ \ref{enthalpy-wrt-pbcn-fm-ls}), supporting
the conclusion that the system is above its magnetic ordering transition
at the room temperature used in the MS experiments\cite{Pasternak-prl99}.
In our calculations, {\it opv}-type Fe$_{2}$O$_{3}$, often considered to be 
another possible candidate structure for the high pressure phase, is 
never the most stable one in this pressure range.

\begin{table*}[t]
\caption{\label{table1}Calculated octahedral parameters
at the  GGA and GGA + U level: bond distortion index, $\delta d_{Fe-O}$,
octahedral angle variance, $\Delta\theta^2$, maximum bond length
(BL$_{max}$), minimum bond length (BL$_{min}$), interaction strength,
$\gamma_{_{NN}}$ and octahedral volume (V$_{oct}$) at the HS $\rightarrow$
LS transition. Three different antiferromagnetic configurations for the
corundum-type phase are named as AF1, AF2 and AF3 and the most stable
ground state magnetic (antiferro) configuration is considered for Rh$_{2}$O$_{3}$(II)
and {\it opv} structures. Overall averages, and their variation (in
parenthesis), for all structural parameters are also indicated. Note 
that GGA + U calculations are done only for structures and magnetic 
configurations that are relevant (having relative stability at 
certain pressures) at the GGA level. And the values of $U$ used here are our 
calculated values for the AF1 corundum type (3.3 eV) and 
FM Rh$_{2}$O$_{3}$(II) type (5.6 eV) Fe$_{2}$O$_{3}$. }
\begin{ruledtabular}
\begin{tabular}{lccccccccc}
 &&  $\delta d_{Fe-O}$ & $\Delta\theta^2$ & BL$_{max}$ & BL$_{min}$ & $\gamma_{_{NN}}$ & V$_{oct}$  \\
 &&  & $(\deg.^{2})$ & (\AA) & (\AA) &  & \\

\hline
{\bf Corundum-type:}\\

\quad AF1({\scriptsize GGA})     && 0.0427 - 0.0248  & 77.8 - 34.0 & 2.012 - 1.912 & 1.848 - 1.820 & 1.360 - 1.422 & 9.28 - 8.56\\
\quad AF2({\scriptsize GGA})     && 0.0298 - 0.0234  & 85.2 - 36.5 & 2.056 - 1.931 & 1.888 - 1.838 & 1.330 - 1.407 & 9.71 - 8.78 \\
\quad AF3({\scriptsize GGA})     && 0.0480 - 0.0246  & 81.0 - 35.8 & 2.036 - 1.936 & 1.917 - 1.848 & 1.311 - 1.401 & 9.96 - 8.91\\ 
\quad AF1({\scriptsize U=3.3 eV})&& 0.0419 - 0.0202 & 107.1 - 44.1 & 1.994 - 1.874 & 1.835 - 1.801 & 1.380 - 1.461 & 8.91 - 8.14 \\
\quad AF1({\scriptsize U=5.6 eV})&& 0.0388 - 0.0196 & 120.0 - 53.2 & 1.984 - 1.864 & 1.838 - 1.794 & 1.382 - 1.470 & 8.81 - 7.99\\ \\
                                        
{\bf Rh$_{2}$O$_{3}$-type:}\\
\quad AF({\scriptsize GGA})&&       0.0375 - 0.0202 & 129.3 - 101.5 & 2.111 - 1.928 & 1.874 - 1.823 & 1.338 - 1.410  & 9.60 - 8.65\\
\quad FM({\scriptsize U=3.3 eV})&&  0.0245 - 0.0184 & 133.0 - 110.3 & 2.096 - 1.935 & 1.947 - 1.825 & 1.290 - 1.401 & 10.04 - 8.55\\
\quad FM({\scriptsize U=5.6 eV})&&  0.0226 - 0.0180 & 134.2 - 111.7 & 2.060 - 1.913 & 1.924 - 1.805 & 1.317 - 1.427  & 9.73 - 8.24\\\\
                                            
{\bf {\it opv}-type:}\\
\quad AF({\scriptsize GGA})&&          0.0363 - 0.0116 & 20.5 - 3.5 & 2.045 - 1.900 & 1.883 - 1.841 & 1.350 - 1.422 & 9.75 - 8.76\\ \\
overall     && 0.035(8) - 0.020(1) & 98.(67) - 58.(95) & 2.04(4) - 1.91(0) & 1.88(4) - 1.82(2) & 1.33(9) - 1.42(4) & 9.5(3) - 8.5(1)\\
%overall (old)     && 0.039(9) - 0.018(7) & 75(55) - 53(50) & 2.03(3)-1.92(2)& 1.88(4) - 1.83(1) & 1.33(3)-1.41(1) & 9.6(3) - 8.7(2)\\
\end{tabular}
\end{ruledtabular}
\end{table*}

All the magnetic configurations, in all investigated structures, show a
spin collapse from $\sim$ 3.8 $\mu_{B}$/Fe atom to $\sim$ 1 $\mu_{B}$
/Fe atom at some critical pressure.  In hematite, the crossover from
the HS to LS occurs at around 20, 25 and 36 GPa for the three different
antiferromagnetic structures consistent with the elemental unit cell in
Fig.\  \ref{structure}(a), while for the ferromagnetic one it occurs at
$\sim$ 5 GPa. This value is $\sim$ 30 GPa for the Rh$_{2}$O$_{3}$(II)
type Fe$_{2}$O$_{3}$ in the most stable magnetic (antiferromagnetic)
configuration.  Also {\it opv}-type Fe$_{2}$O$_{3}$ displays a sharp
spin transition as a function of pressure. In this structure there
are two types of polyhedral cages (with 6-fold and 8-fold coordination)
surrounding the cations and while Fe ions in 8-fold coordinated polyhedra
loose their spin gradually, Fe ions in the octahedral cages display a
sudden drop from HS to LS at $\sim$ 21 GPa. In all cases a volume collapse
of about $\sim$ 7 -- 10 \% at the spin transition is found, consistent
with a first order transition and in agreement with experimental evidence.
Hence, regardless of the considered crystal structures, the Fe atoms
sitting in octahedral cages display a HS to LS transition under pressure.

From the experimental point of view it is still controversial whether the
spin transition occurs in the high-pressure or low-pressure structural
phase. We believe that this spread in experimental results reflects
the fact that spin transition is a general phenomenon occurring in all
relevant structures and with a simple common origin as shown below.

\section{LOCAL STRUCTURE}
In order to clarify the interplay between spin and structural
transitions, we want to identify a few structural indicators that could
be used to characterize the HS to LS transition and, as in previous
studies,\cite{Takemoto-ic73,Gutlich-sb81,Konig-pic87,Konno-bcsj91,
Moliner-ica99,Guionneau-jmc99, Marchivie-ac03} we will concentrate on the
Fe coordination sphere. Pressure and/or temperature induced (anisotropic)
strain causes deformation of the octahedra, which can be quantified by
parameters such as the Fe--O bond-length distortion, \(\delta d_{Fe-O}
= \frac{1}{6D} \sum^{6}_{i=1} |\Delta d_{i}|\), where D is the average
octahedral Fe--O bond-length and $\Delta d_{i}$ is the deviation of
the {\it i-th} Fe--O bond-length from the average, and the octahedral
angle variance,\cite{Robinson-science71} given by \(\Delta \theta^2 =
\sum^{12}_{j=1} (\theta_{j} - 90)^{2}/11\), where \(\theta_{j}\) is the
{\it j-th} O--Fe--O angle whose ideal value is 90$^\circ$.

In all structures, the octahedra become more regular at the transition
as can be inferred from the sudden jump of both parameters,
from higher to lower values, reported in Table \ref{table1}. However,
\(\delta d_{Fe-O}\) and \(\Delta \theta^2\) display very different values
for different structures and hence these parameters alone are not suitable
as a quantitative fingerprint of the occurrence of the spin transition.

Other octahedral structural parameters can be defined such as the
maximum (BL$_{max}$) and minimum (BL$_{min}$) Fe--O bond-lengths,
the octahedral volume (V$_{oct}$) and a parameter measuring the
nearest-neighbor interaction, defined as \(\gamma_{_{NN}} = \sum^{6}_{j=1}
e^{-d_{j}/r_O}\), where $r_O$ = 1.3 \AA \ is of the order of the Oxygen
ionic radius and \(d_{j}\)'s are the Fe--O bond lengths.  As can be seen
in Table \ref{table1}, BL$_{max}$, BL$_{min}$ and V$_{oct}$ show jumps to
lower values across the transition. Moreover they present very similar
values in all structures, both just before but especially just after the
transition when the spread of these parameters in different structures
is of the order of a few percent. Similarly, the interaction parameter,
$\gamma_{_{NN}}$, also show a jump, this time from a lower to a larger
value, again with a spread of a few percent.

It is worth stressing that, in spite of the fact that the HS to LS
transition occurs at different pressures in different structures, and the
results accumulated contain both GGA and GGA +$U$ calculations, for all
the structural parameters considered here, the spread in their calculated
values, just before and especially just after the transition, is smaller
than the jump they experience across the transition, thus revealing the
common mechanism operating in the different structures. All the above
three structural parameters are therefore good indicators that could be
used to monitor the approaching of a system to its critical pressure.

\begin{figure}[!tbp]
\includegraphics[scale=0.29]{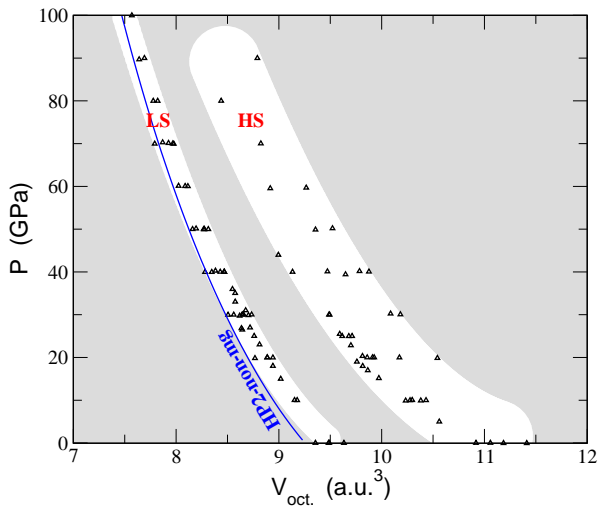}
\caption{(Color online) Octahedral volumes for the HS and LS states. The data include
corundum-type, Rh$_{2}$O$_{3}$-type and {\it opv}-type structures at
the GGA level as well as for corundum-type and Rh$_{2}$O$_{3}$-type
structures at the GGA+U level. }

\label{vol-collapse}
\end{figure}

The octahedral volume is possibly the most fundamental one
and in Fig. \ref{vol-collapse} we report the evolution under
pressure of the octahedral volumes of the different structural
phases in the HS and LS states. The data include corundum-type,
Rh$_{2}$O$_{3}$(II)-type and {\it opv}-type in the GGA scheme and,
for corundum-type and Rh$_{2}$O$_{3}$(II)-type, in the GGA+U scheme
($U$=3.3 and 5.6 eV, respectively, as calculated according to Ref.\
\onlinecite{Cococcioni-prb05}). The transition pressures\cite{note-on
HS to LS with GGA+U} range from $\sim$ 0 GPa to $\sim$ 70 GPa and the
crystal packing is different for different structures. In spite of all
the above differences, the HS and LS octahedral volumes follow two well
defined equations of state. The gap between these two equations of state
corresponds to the octahedral volume collapse at the transition. This gap
is larger at lower pressures and slowly decreases with pressure. The
finite width of these two equation of states can be attributed to
the fact that different structure display different inter-polyhedral
surroundings, which are known to play some role in controlling the
spin transition,\cite{Guionneau-jmc99, Marchivie-ac03} as well as
the fact that the results from different DFT scheme (GGA and GGA+U)
are collected together. Yet, it is remarkable that the finite width
in the equation of states for the octahedral volume is only about 9\%
for the HS state and just $\sim$ 3\% in the LS one.
\par
A possible explanation for the larger uncertainty displayed in the
HS equation of state lays in the fact that in this state, owing to
larger Fe--O bonds and angular distortions (see table \ref{table1}), the
influence of the surroundings on the octahedra is more significant thus
inducing larger spread in the octahedral volume. On the contrary, in the
denser LS phase the octahedra become more regular (table \ref{table1})
and the increased Fe--O bond strength, $\gamma_{_{NN}}$, reduces the
relative influence of the surroundings, thus contributing to a smaller
spread in the equation of state.

Based on the above two regions in Fig. \ref{vol-collapse} it might
also be possible to estimate the amount of volume collapse or the spin
transition pressure at 0K for similar kind of materials, provided any
one of the quantities is known. Temperature effects on the transition
pressure are however dramatic and are presently beyond our understanding.

\section{DISCUSSION}

Given a reliable and consistent structural parameter to characterize the
spin transition, let us now turn to an analysis of the main experimental
facts involved in its interplay with the structural one. The co-existence
of both low- and high-pressure structural phases over a finite width of
pressure/temperature has already been pointed out by a number of studies
\cite{Pasternak-prl99, Rozenberg-prb02,Shim-am02}.
Some controversy still exists \cite{Pasternak-prl99,Badro-prl02}
on the nature of the intervening phases. The resistivity profile in
[\onlinecite{Pasternak-prl99}] clearly identifies three pressure regions:
a magnetic insulating phase at lower pressure, a nonmagnetic metallic
phase at very high pressure and a mixture of them in between. In
particular the nonmagnetic nature of the metallic phase in the
intermediate region was assigned on the basis of the assignment of
a doublet in the M\"ossbauer spectra as arising from nonmagnetic
Fe atoms. We notice however that this assignment is not completely
compelling since also a para-magnetic phase, above its ordering
transition-temperature, would contribute a doublet in MS. This
interpretation would be consistent with the small energy spread between
the different ferromagnetic configurations found in our calculations
and reported in the inset of Fig.~\ref{enthalpy-wrt-pbcn-fm-ls}. Indeed,
the x-ray K$\beta$ emission spectra in [\onlinecite{Badro-prl02}] showed
evidence of (para-)magnetism, although very weak. This study did not
address the conductivity of the various phases.

The picture arising from our calculations agrees with all the experimental
evidence and helps clarifying the situation. As can be seen from
Fig. \ref{enthalpy-wrt-pbcn-fm-ls}, the system has three broad regions in
terms of magnetic states: a HS region at low pressure ({\it R-I}), a LS,
but still (para-)magnetic, phase at intermediate pressure ({\it R-II})
and a nonmagnetic phase at very high pressure({\it R-III}). While the HS
phase is insulating, both corundum-type and Rh$_{2}$O$_{3}$(II)-type LS
phases as well as the nonmagnetic phases are metallic. The HS insulating
phase is followed by the LS metallic phase in either of the structural
phases\cite{note-on-HP1-metallic-phase}, HP1 or HP2, and we believe
it is not constrained by any sharp boundary. In this crossover region,
for a finite width on the pressure scale, the HS $\rightleftharpoons$
LS transformation occurs in any particular octahedron depending on
the volume that the particular octahedron acquires in that condition.
Our explanation for the spin transition goes as follows: for a particular
octahedron, the Fe--O bond lengths and the octahedral volume decreases
gradually with increasing pressure; at a certain point in the pressure
scale, when the minimum Fe--O bond length (see Table \ref{table1}) in that
particular octahedron reaches a value for which the octahedral crystal
field dominates over the exchange energy, that particular octahedron
acquires a new spin state switching from its $^{3}t_{2g}$ $^{2}e_{g}$
state to $^{5}t_{2g}$ $^{0}e_{g}$, resulting in the collapse of the
octahedral volume. Similar octahedra with equivalent surroundings will
experience this transition at the same pressure. While the transition
pressure will vary for structures with different crystal packing.

Our observation that HS $\rightleftharpoons$ LS transformation depends on
a local property, the octahedral volume, is in agreement with and explains the
observed sluggishness of the structural transition\cite{Rozenberg-prb02,
Shim-am02}. Inhomogeneity in the experimental pressure and temperature
fields, the presence of defects and impurities in the experimental samples
locally affect the octahedral volume around Fe ions and therefore
upon increasing pressure the HS to LS transition will occur at different
pressures in different octahedra depending on when the critical octahedral
volume we have identified is reached. The local volume-collapse associated
with spin transition will contribute to enhance the inhomogeneity of the
pressure-field in the sample and the transition will proceed gradually
in a nucleation-and-growth way.

In the same pressure range also the structural transition occurs. Minor
crystallographic changes, brought about by the alteration of one
of the coordinating oxygen atoms, lead to distorted corundum type
or Rh$_{2}$O$_{3}$(II) type from the corundum type structure. An
important issue is whether spin and structural transitions are
strongly inter-dependent or not. Our results indicate that the two
type of transitions are not strongly dependent, rather the HS to
LS transition is a general phenomenon occurring in all structural
phases. This conclusion is in agreement with experimental evidence.
Badro {\it et. al.}\cite{Badro-prl02} has indeed shown that the spin
transition is not necessary for the structural transition to occur,
and vice versa. Hence, based on our results and experimental findings,
one can qualitatively conclude that the spin and structural transitions
are independent phenomena, in the case of Fe$_{2}$O$_{3}$, at least.

The HS $\rightarrow$ LS and HP1 $\rightarrow$ HP2 transition are completed
by the upper bound of the intermediate region ({\it R-II}), which then
progresses slowly to the nonmagnetic HP2 phase.  Our calculated boundary
for LS HP2 $\rightarrow$ nonmagnetic HP2 is $\sim$ 120 GPa at zero
temperature, which is rather far from the room temperature experimental
value of $\sim$ 72 GPa [\onlinecite{Pasternak-prl99}]. Even assuming an,
expected, strong temperature effect the discrepancy remains large. Still,
accepting the inadequacy of DFT in predicting the numerical value of
the transition pressures, we stress that DFT at the GGA level predicts
correctly the sequence and details of the transitions.
Taken together, our DFT results and the two main experimental works
\cite{Pasternak-prl99, Badro-prl02} provide a clearer picture of the
pressure evolution of the system across three regions: a HS magnetic
phase ({\it R-I}), a LS, but still (para-)magnetic, phase at intermediate
pressure ({\it R-II}) and a nonmagnetic phase at very high pressure({\it
R-III}).

\begin{figure}[!ht]
\includegraphics[scale=0.28]{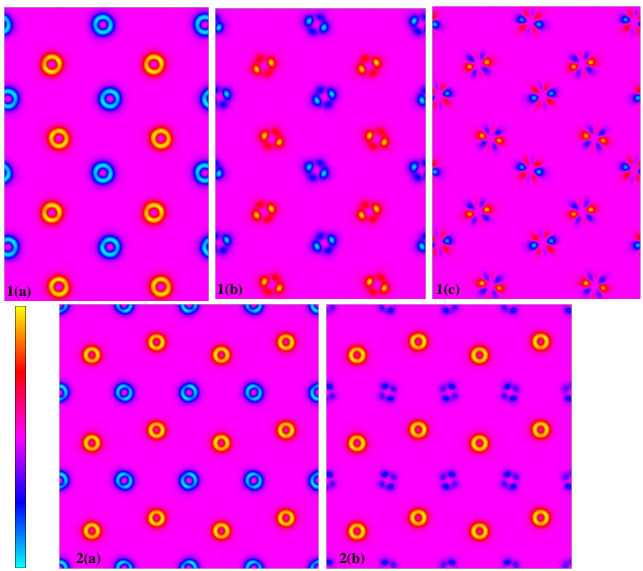}
\caption{ (Color online) Spin-density difference plots for the HP2  and {\em
opv} across the HS $\rightarrow$ LS transition. Upper panels, 1(a), 1(b)
and 1(c), refers the spin-density difference to HP2 in the {\it R-I},
{\it R-II}, and {\it R-III} regions, respectively. Lower panels, 2(a)
and 2(b), display the spin-density difference for {\em opv} before and
after the HS $\rightarrow$ LS transition, respectively. For the HP2
structure the plot is for (010) plane at 0.25 interplanar spacing away
from the origin. For {\em opv} (100) plane at half of the interplanar
spacing from the origin is shown. Note the spherical nature of the
spin-density difference for Fe inside the 8-fold coordinated cage after
the transition. The planes are antiferromagnetic, as can be seen in the
color differences. For convenience the scale is shown in the left and two
extreme points correspond to maximum values in the positive and negative
direction.  }

\label{sd-pbcn-pbnm}
\end{figure}

To further elucidate the evolution of the system across the spin
transition and display the common mechanism that drives it in different
crystal structures we present in Fig. \ref{sd-pbcn-pbnm} spin-density
difference plots for HP2 (upper panels) and {\em opv} (lower panels).
Before the spin transition (see 1(a) and 2(a)) all Fe atoms are
characterized by spherical spin-density difference contour plots,
corresponding to the fact that the level occupation is $^{3}t_{2g}$
$^{2}e_{g}$ for HS.  Just after the spin transition (see 1(b) and 2(b))
Fe atoms in octahedral coordination sites switch to the LS state. Their
spin-density difference is characterized by four-lobe, due to the local
$^{5}t_{2g}$ $^{0}e_{g}$ level occupation. Notice that in panel 2(b) only
the spin-density difference of the octahedral sites has changed nature;
Fe atoms in the 8-fold coordinated sites remain HS. Finally panel 1(c)
shows the spin density difference plot of HP2 when the system looses its
spin completely and becomes nonmagnetic. On each site the averaged spin
density vanish. The increased proximity of Fe and O atoms with increasing
pressure enhances further the splitting of the $d$ orbitals along with
the pressure induced broadening of the states and a simultaneous transfer
of some charge from O--$p$ to Fe--$d$. This eventually leads to a spin
state characterized by $^{6}t_{2g}$ $^{0}e_{g}$ level occupation.
The similarity in the spin-density difference obtained in corresponding
sites in different structures clearly show the common mechanism at work.

\section {CONCLUSIONS}

In conclusion, we have shown that, in spite of the generally
believed intricate relations with different structural parameters,
the phenomenon of spin transition can well be described by a rather
simple local structural parameter, {\it i.e.}, the octahedral volume, at
least in Fe$_{2}$O$_{3}$, if not in general. The HS and LS octahedral
volumes, separated by a well defined volume gap, are governed by two
distinct equations of state. For different structural phases, density
functional theory, both at the GGA and GGA+U levels, reproduces this spin
transition at a critical value of the octahedral volume.  Analysis of
our calculations suggests that at intermediate pressures the system is
in a LS paramagnetic phase, thus clarifying some experimental controversy.

\acknowledgements
Calculations were performed at SISSA and at the CINECA computing center,
also thanks to INFM computing grants.

\clearpage


\begin{thebibliography}{99}

%\bibitem{general}
\bibitem{Takemoto-ic73}
J.H.\ Takemoto, and B.\ Hutchinson, Inorg.\ Chem.\ {\bf 12}, 705 (1973);

\bibitem{Gutlich-sb81}
P.\ G\"utlich,  Struct.\ Bonding,\ {\bf 44}, 83 (1981);

\bibitem{Konig-pic87}
E.\ K\"onig, Prog.\ Inorg.\ Chem.\ {\bf 35}, 527 (1987);

\bibitem{Konno-bcsj91}
M.\ Konno, and M.\ Mikami-kido, Bull.\ Chem.\ Soc.\ Jpn.\ {\bf 64}, 339 (1991);

\bibitem{Moliner-ica99}
N.\ Moliner, M.C.\ Munoz, S.\ Letard, J.-F.\ Letard, X.\ Solans, R.\ Burriel, M.\ Castro, O.\ Kahn, and J.-A.\ Real, 
Inorg.\ Chem.\ Acta {\bf 291}, 279 (1999).

\bibitem{Guionneau-jmc99}
P.\ Guionneau, J.-F.\ L\,etard, D.S.\ Yufit, D.\ Chasseau, G.\ Bravic, A.E.\ Goeta, J.A.K.\ Howard, and O.\ Kahn, 
J.\ Mater.\ Chem.\ {\bf 9}, 985 (1999).

\bibitem{Marchivie-ac03}
M.\ Marchivie, P.\ Guionneau, J.-F.\ L\,etard and D.\ Chasseau, Acta.\
Cryst.\ B{\bf 59}, 479 (2003).

\bibitem{Hubbard64}
J.\ Hubbard, Proc.\ R.\ Soc.\ London A {\bf 277}, 237 (1964).

\bibitem{Morin-pr50}
F.J.\ Morin, Phys.\ Rev.\ {\bf 78}, 819 (1950).

\bibitem{Searle-prb70}
C.W.\ Searle and G.W.\ Dean, Phys.\ Rev.\ B {\bf 1}, 4337 (1970).

\bibitem{Levinson-prb71}
L.M.\ Levinson, Phys.\ Rev.\ B {\bf 3}, 3965 (1971).

\bibitem{Chow-prb74}
H.\ Chow and F.\ Keffer, Phys.\ Rev.\ B {\bf 10}, 243 (1974).

\bibitem{Sandratskii-epl96}
L.M.\ Sandratskii and J.\ Ku\"bler, Europhys.\ Lett.\ {\bf 33}, 447 (1996).

\bibitem{Reid-jgr69}
A.F.\ Reid and A.E.\ Ringwood, J.\ Geophys.\ Res.\ {\bf 74}, 3238 (1969).

\bibitem{Mcqueen-gsa66}
R.G.\ McQueen and S.P.\ Marsh, in Handbook in Physical Constants, edited
by S.P.\ Clark, Memoir 97 of the Geological Society of America, Inc.\
(Geological Society of America, New York, 1966), revised ed., p.~153.

\bibitem{Shannon-jssc70}
R.D.\ Shannon and C.T.\ Prewitt, J.\ Solid State Chem.\ {\bf 2}, 134 (1970).

\bibitem{Yagi-capj82}
T.\ Yagi and S.\ Akimoto, in High Pressure Research in Geophysics,
edited by S.\ Akimoto and M.H.\ Manghnani (Center Acad.\ Publ.\ Japan,
Tokyo, 1982), p.~81.

\bibitem{Suzuki-ktksp85}
T.\ Suzuki, T.\ Yagi, A.\ Akimoto, A.\ Ito, S.\ Morimoto, and S.\ Syono, 
in Solid State Physics under Pressure, edited by S.\ Minomura
(KTK Scientific Publishers, Tokyo, 1985), p.~149.

\bibitem{Nasu-hi86}
S.\ Nasu, K.\ Kurimoto, S.\ Nagatomo, S.\ Endo, and F.E.\ Fujita,
Hyperfine Interact.\ {\bf 29}, 1583 (1986).

\bibitem{Syono-ssc84}
Y.\ Syono, A.\ Ito, S.\ Morimoto, S.\ Suzuki, T.\ Yagi, and S.\ Akimoto, 
Solid State Commun.\ {\bf 50}, 97 (1984).

\bibitem{Olsen-ps91}
J.\ Staun~Olsen, C.S G.\ Cousins, L.\ Gerward, H.\ Jhans, and B.J.\ Sheldon, 
Phys.\ Scr.\ {\bf 43}, 327 (1991).

\bibitem{Pasternak-prl99}
M.P.\ Pasternak, G.Kh.\ Rozenberg, G.Y.\ Machavariani, O.\ Naaman, R.D.\ Taylor, and R.\ Jeanloz, 
Phys.\ Rev.\ Lett.\ {\bf 82}, 4663 (1999).

\bibitem{Badro-prl02}
J.\ Badro, G.\ Fiquet, V.V.\ Struzhkin, M.\ Somayazulu, H.-K.\ Mao, G.\ Shen, and T.\ LeBihan, 
Phys.\ Rev.\ Lett.\ {\bf 89}, 205504 (2002).

\bibitem{Shim-am02}
S.-H.\ Shim and T.S.\ Duffy, Am.\ Miner.\ {\bf 87}, 318 (2002).

\bibitem{Rozenberg-prb02}
G.Kh.\ Rozenberg, L.S.\ Dubrovinsky, M.P.\ Pasternak, O.\ Naaman, T.\ Le Bihan, and R.\ Ahuja, 
Phys.\ Rev.\ B {\bf 65}, 064112 (2002).

\bibitem{Ono-jpcs04}
S.\ Ono, T.\ Kikegawa, and Y.\ Ohishi, J.\ Phys.\ Chem.\ Solids {\bf 65}, 1527 (2004);

\bibitem{Ono-jpcm05}
S.\ Ono, K. Funakoshi, Y. Ohishi, and E. Takahashi, J.\ Phys.:\ Condens. Matter {\bf 17}, 269 (2005).

\bibitem{Wentzcovitch-prb91}
R.M.\ Wentzcovitch, Phys.\ Rev.\ B {\bf 44}, 2358 (1991). 

\bibitem{Vanderbilt-prb90}
D.\ Vanderbilt, Phys.\ Rev.\ B {\bf 41}, 7892 (1990).

\bibitem{pseudo} 
We used the pseudopotentials Fe.pbe-nd-rrkjus.UPF
and O.pbe-rrkjus.UPF from the {\sc Quantum-ESPRESSO} distribution.
See http://www.quantum-espresso.org

\bibitem{Perdew-prl96}
J.P.\ Perdew, K.\ Burke, and M.\ Ernzerhof, Phys.\ Rev.\ Lett.\ {\bf 77}, 3865 (1996). 

\bibitem{Cococcioni-prb05}
M.\ Cococcioni and S.\ de~Gironcoli, Phys.\ Rev.\ B {\bf 71}, 035105 (2005).

\bibitem{QE}
{\sc Quantum-ESPRESSO} is a community project for high-quality
quantum-simulation software, based on density-functional theory, and
coordinated by Paolo Giannozzi. See http://www.quantum-espresso.org.

\bibitem{Robinson-science71}
K.\ Robinson, G.V.\ Gibbs, and P.H.\ Ribbe, Science {\bf 172}, 567 (1971).

\bibitem{note-on HS to LS with GGA+U}
In Ref.[\onlinecite{Rollmann-prb04}] no spin transition was found within
GGA+U in the corundum-type phase. Our calculations, however, do find
an HS to LS transition even in this structure, although at an higher
pressure than in GGA.

\bibitem{Rollmann-prb04}
G.\ Rollmann, A.\ Rohrbach, P.\ Entel, and J.\ Hafner, Phys. Rev. B {\bf 69}, 165107 (2004).

\bibitem{note-on-HP1-metallic-phase}
LS HP1 metallic phase is only metastable in the stability region of LS HP2 metallic phase. Also our calculations doesn't favor the nonmagnetic HP1 phase with respect to the LS HP1 phase upto about 150 GPa. 

\end{thebibliography}
\end{document}